\documentclass[aps,prd,showpacs,nofootinbib,floats,floatfix,preprintnumbers,groupedaddress,twocolumn]{revtex4-1}
\usepackage{graphicx,epsfig}
\usepackage{dcolumn}
\usepackage{bm}
\usepackage{latexsym}
\usepackage[table]{xcolor}
\usepackage{booktabs}
\usepackage{color}
\usepackage{tabularx}
\usepackage{ulem}
\usepackage{hyperref}
\usepackage{float}
\usepackage{tabularx}
\usepackage{color}
\usepackage{comment}
\usepackage{physics}
\usepackage{subfigure}
\usepackage{tikz}
\usepackage{hyperref}
\usepackage{colortbl}
\usepackage{listings}
\usepackage{amsmath,amsfonts,amssymb}
\usepackage{fancyhdr}
\usepackage{orcidlink}
\usepackage{hyperref}
\usepackage{natbib}
\usepackage{multirow}
\hypersetup{
	colorlinks   = true, 
	urlcolor     = blue, 
	linkcolor    = blue, 
	citecolor    = red    
}

\hypersetup{
	colorlinks   = true, 
	urlcolor     = blue, 
	linkcolor    = blue, 
	citecolor    = red 
}

\definecolor{lightblue}{RGB}{173,216,230}
\definecolor{lightgreen}{RGB}{144,238,144}
\definecolor{magenta}{RGB}{255,0,255}
\definecolor{myolive}{RGB}{128,128,0}
\definecolor{masteredyellow}{RGB}{255,255,102}
\definecolor{mymaroon}{RGB}{128,0,0}

\begin{document}

	\title{Probing Chaos in Schwarzschild-de Sitter Spacetime: The Role of Black Hole and Cosmological Horizons}

	\author{Surojit Dalui$^{1}$\orcidlink{0000-0003-1003-8451}}
        \email{surojitdalui@shu.edu.cn,\\
        surojitdalui003@gmail.com}

        \author{Soumya Bhattacharya$^{2,3}$\orcidlink{0000-0003-2540-7504}}
        \email{soumya557@gmail.com}

\author{Chiranjeeb Singha $^3$ \orcidlink{0000-0003-0441-318X}}
\email{chiranjeeb.singha@iucaa.in}

        \affiliation{ $^{1}$Department of Physics, {Shanghai University}, 99 Shangda Road, Baoshan District, Shanghai 200444, China\\
        	$^{2}$Department of Astrophysics and High energy physics,  S.N. Bose National Center for Basic Sciences, Kolkata 700106, India\\
        	$^3$Inter-University Centre for Astronomy and Astrophysics , Post Bag 4, Ganeshkhind, Pune, 411 007, India }


\begin{abstract}
   In this paper, we study the motion of a massless, chargeless particle in Schwarzschild-de Sitter spacetime, revealing exponential radial growth and potential chaos in an integrable system. Poincaré sections show regular Kolmogorov-Arnold-Moser (KAM) tori when black hole and cosmological horizons are distant, but distortions and chaos emerge as they converge. As the horizons coincide, the Poincaré sections fully contract and vanish, marking the system's transition to Nariai spacetime. \textit{Our analysis also suggests that, within the parameter range explored, the event horizon exerts a comparatively more substantial chaotic influence on the system, primarily due to its consistent proximity.} Additionally, we analyze the Lyapunov exponents to quantify the degree of chaos in the system. Our findings indicate that as the closeness of the two horizons increases, the most prominent Lyapunov exponent also increases, signifying a rise in chaotic behavior. By examining the long-term saturation values of the Lyapunov exponents, we confirm that they consistently comply with the Maldacena-Shenker-Stanford (MSS) bound.
\end{abstract}
	

	\maketitle

\noindent
\section{Introduction}
The study of particle dynamics around massive objects is a central focus in physics. The general theory of relativity (GR) provides a successful framework for explaining various phenomena related to the motion of astrophysical objects, including the bending of light caused by the presence of mass in spacetime. Among compact objects, black holes remain some of the most fascinating entities in the Universe. Theoretically, black holes are solutions to Einstein’s equations of motion, characterized by a boundary known as the event horizon, a one-way membrane through which nothing can escape, according to classical physics. The recent detection of gravitational waves by LIGO \cite{LIGOScientific:2016aoc, LIGOScientific:2017vwq} and event horizon telescope (EHT) \cite{EHT:1, EHT:2, EHT:3, EHT:4, EHT:5}  has confirmed that black holes are not merely theoretical constructs; they indeed exist in spacetime. Now, researchers continue to focus extensively on understanding the physics of black holes and the diverse phenomena they generate, both at astrophysical and quantum scales.

Over the past three decades, there has been growing interest in de Sitter (dS) space at both astrophysical and quantum scales. This attention has been driven by two primary factors: Observational evidence suggesting that the universe may currently be transitioning toward an asymptotic dS state, evolving into a pure dS configuration.
The success of the AdS/CFT correspondence \cite{Maldacena_1999} has sparked extensive research into quantum gravity in dS space and efforts to develop a comparable dS/CFT framework \cite{Strominger_2001}. While significant progress has been made in understanding dS space, much remains unknown about Schwarzschild-de Sitter (SdS) space-time. The key difference between SdS and dS (or Schwarzschild) space-time lies in the presence of multiple horizons. SdS space-time has both a cosmological horizon and an event horizon, whereas dS and Schwarzschild space-time each possess only one horizon. This dual-horizon structure of SdS space-time introduces new complexities in the study of particle dynamics, as the interplay between these horizons can lead to novel physical phenomena. Unlike Schwarzschild spacetime, where the event horizon is the dominant causal boundary, in SdS spacetime, the presence of a second horizon—the cosmological horizon—can significantly influence the motion of test particles. This raises several intriguing questions: How does the cosmological horizon impact in particle dynamics? Does the presence of multiple horizons alter the stability of geodesics in a fundamental way? Addressing these questions is crucial for developing a more complete understanding of black hole and cosmological spacetimes.

Recently, the motion of a massless, chargeless particle in Schwarzschild's black hole spacetime has been studied \cite{Dalui_2019}. It was demonstrated that the radial motion exhibits exponential growth, indicating the potential onset of chaos in the dynamics of an otherwise integrable system influenced by the presence of a black hole horizon. Poincaré sections of the trajectories were analyzed to verify this behavior, with the particle's motion confined using a harmonic trap. Subsequently various other directions within Einstein gravity have been explored in \cite{Dalui_2020, Dalui_2020_0, Dalui_2020_1, Dalui_2022}. Similar studies have also been conducted for black holes beyond Einstein's gravity \cite{Bera_2022, Das_2024}. However, no similar analysis has been conducted for scenarios where a cosmological horizon is present instead of a black hole horizon. This raises the question of whether the cosmological horizon affects particle dynamics in a similar or distinct manner. Investigating whether chaos probes can serve as a tool to determine the location of the cosmological horizon would be particularly intriguing.

In this paper, we present a detailed analysis of the motion of a massless, chargeless particle in Schwarzschild-de Sitter spacetime. Our study is motivated by the need to explore the chaotic properties of trajectories in the presence of multiple horizons and to determine whether the interplay between the black hole and cosmological horizons fundamentally alters the onset of chaos. Our analysis reveals that the radial motion shows exponential growth, suggesting the potential onset of chaos in the dynamics of an otherwise integrable system influenced by the presence of horizons. To confirm this behavior, we examine Poincaré sections and Lyapunov exponents of the trajectories, confining the particle's motion using a harmonic trap.


The results clearly illustrate the influence of both the black hole and cosmological horizons on the onset of chaos and the transition to fully chaotic behavior within the specified range of the parameter \( y \) (which characterizes the relative separation of the horizons). Moreover, in this analysis, we establish that chaos is inevitable in the presence of horizons and that the Lyapunov exponent has an upper bound. Remarkably, this bound can be expressed as 
$\lambda_L \leq \kappa,
$ where $\kappa$ represents the effective surface gravity of black hole and cosmological horizons. Surface gravity, which corresponds to the acceleration experienced by a particle as measured by a distant observer near the horizon, was recently predicted \cite{Hashimoto_2017} in a completely different analysis involving a massive particle. Furthermore, we found that this value aligns with the bound predicted in \cite{C_ceres_2023} by studying out-of-time-order correlators of observables in the Sachdev-Ye-Kitaev (SYK) model. This demonstrates that the bound has a universal character.

Let us now consider the predictions that can be drawn from our current analysis. Here, we focus exclusively on massless particles following outgoing trajectories, where the radial trajectory corresponds to a radial null geodesic. Notably, this same null geodesic is responsible for the Hawking radiation of particles from the horizons (see \cite{Parikh_2000} for an understanding of Hawking radiation as a tunneling process). An important aspect of our analysis is that the radiated particles, after escaping the black hole horizon, do not exhibit purely regular motion but instead undergo chaotic dynamics within the region enclosed by the cosmological horizon. This chaotic behavior arises due to the intricate interplay of gravitational effects from both the horizons and external perturbations from various objects in the Universe. The presence of chaos in the motion of these particles suggests that horizons not only act as sources of Hawking radiation but also significantly influence the post-emission dynamics of radiated particles. This raises intriguing questions regarding the observational signatures of such chaotic behavior, particularly in the context of black hole thermodynamics and cosmological evolution. If chaos probes can effectively capture the underlying instability induced by the horizons, they might offer new avenues for understanding the interplay between quantum and classical aspects of horizon physics.


The paper is organized as follows: in section \ref{sec2}, we briefly review the line element and horizons for Schwarzschild de Sitter (SdS) spacetime. In section \ref{sec3}, we consider a massless particle in sds spacetime and study the dynamics of the particle. In section \ref{sec4}, we first study the Poincar$\Acute{e}$ sections of the particle trajectories. Then, we investigate the Lyapunov exponents to quantify the chaos observed in the Poincar$\Acute{e}$ sections. 
Finally, in Section \ref{conclusion}, we present a summary of our findings and discuss possible avenues for future research.  

\textbf{\emph{Notations and Conventions}}:  
Throughout this paper, we employ the positive signature convention, where the Minkowski metric in \(3+1\) dimensions is expressed in Cartesian coordinates as \(\text{diag}(-1, +1, +1, +1)\). Additionally, we work in natural units, setting \(G = c = 1\).

\section{Schwarzshild de Sitter spacetime} \label{sec2}

\begin{figure}[H]
	\begin{center}
		\includegraphics[width=1.0\linewidth]{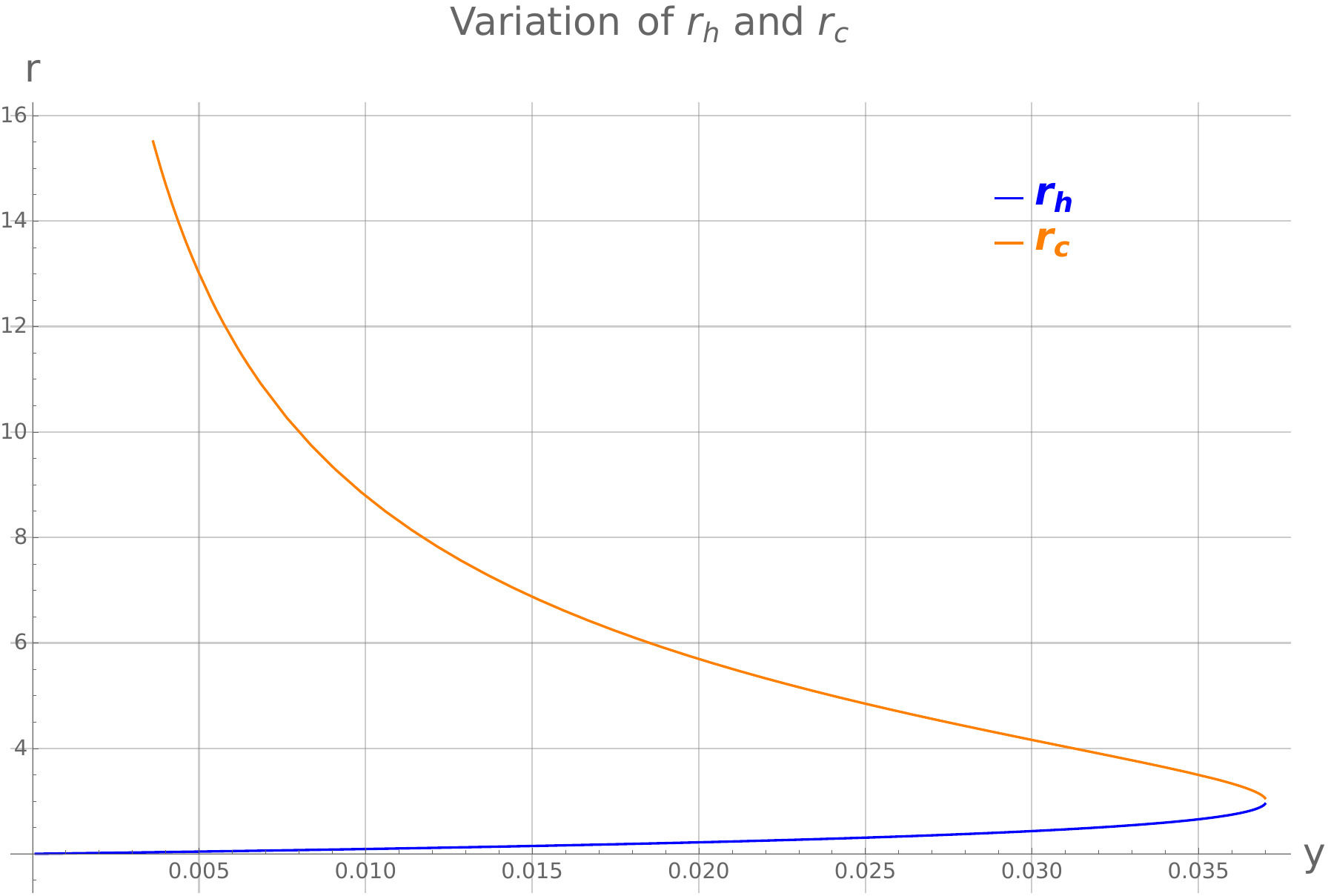}\label{plot1}
	\end{center}
	\caption{This figure shows the variation of $r_{h}$ (blue) and $r_{c}$ (orange) with respect to the variation of $y$ and one can see that both the plots approach at the same value to coincide near $y=1/27$ which is known as Nariai space-time.}
	\label{fig}
\end{figure}

In this section, we first briefly discuss about the Schwarzschild de Sitter spacetime. The metric for this spacetime is given below,
\begin{equation}
	ds^2=-g(r)dt^2+\frac{dr^2}{g(r)}+r^2 d\Omega^2~,\label{metric}
\end{equation}
where
\begin{eqnarray}
	g(r)=\left(1-\frac{2M}{r}-\frac{r^{2}}{l^{2}}\right)~.
\end{eqnarray}
Here $M$ is the mass of the black hole, and $l^2$ is related to the positive cosmological constant. The space-time has more than one horizon if $0<y<1/27$ where $y=M^2 /l^2$. The black hole horizon $(r_h )$ and the cosmological horizon $(r_c )$ are located, respectively, at \cite{Shankaranarayanan_2003}
\begin{eqnarray}
	r_{h}&=&\frac{2M}{\sqrt{3y}}\cos{\frac{\pi+\psi}{3}}, \label{rh}\\
	r_{c}&=&\frac{2M}{\sqrt{3y}}\cos{\frac{\pi-\psi}{3}}, \label{rc}
\end{eqnarray}
where
\begin{eqnarray}
	\psi=\cos^{-1}{3\sqrt{3y}}. \label{psi}
\end{eqnarray}
In the limit of $y\rightarrow0$, we get $r_h \rightarrow 2M$ and $r_c \rightarrow l$. (Note  that $r_h <r_c$, i.e., the event horizon is the smallest positive root.) The space-time is dynamic for $r<r_h$ and $r>r_c$. In the limit $y\rightarrow 1/27$, the two horizons event and cosmological coincide and are the well-known Nariai space-time. If $y >1/27$, the space-time is dynamic for all $r>0$. In Fig. (\ref{fig}), we plot The black hole horizon and cosmological horizon for different values of $y$

\section{Dynamics of a massless particle in the SdS background} \label{sec3}
Now the metric (\ref{metric}) admits a timelike Killing vector $\zeta^{\alpha}=(1,0,0,0)$, so that the conserved energy is given by $E=-\zeta^{\alpha}p_{\alpha}$, where $p_{\alpha}$ is the particle's four momentum. Next, to find the energy of the particle in terms of other components of the four-momentum, we use the covariant form of the dispersion relation given by
\begin{equation}
	g^{\alpha\beta}p_{\alpha}p_{\beta}=-m^2 c^2,\label{disp}
\end{equation}
with $m$ being the mass of the particle.

Now, using the dispersion relation Eq. \eqref{disp} (with $c=1$) for the metric Eq. \eqref{metric}, we can get the energy of a massless particle (for $m=0$) as follows (considering the +ve solution of $E$):
\begin{eqnarray}
	E=\sqrt{g(r)\Big[g(r)p_r^2+\dfrac{p_{\theta}^2}{r^2}\Big]}.\label{a2}
\end{eqnarray}
Here, we have assumed that the particle is moving only along the radial $r$ and the angular $\theta$ directions; i.e., the motion of the particle is in the poloidal plane with $p_{\phi}=0$.

Let us now construct the dynamical equations of motion of a probe massless particle in the background geometries arising in the Schwarzschild–de Sitter (SdS) space-time.  Here, the SdS black hole solution has two horizons as we mentioned earlier ($r_h$ and $r_c$) and in the limit $y\rightarrow 1/27$, the two horizons - event and cosmological coincide. So, in this paper, our motivation is to study how both horizons affect the particle's trajectories if our system resides in between the two horizons. So, here we introduce an integrable system probing our massless particle in a harmonic potential along radial  $r$ and angular $\theta$ directions with the oscillator strength $K_r$ and $K_{\theta}$ respectively. By adjusting these strengths, we can confine the particle trajectories in any finite region. The inclusion of harmonic oscillators in the radial and angular directions is a choice of the external potential rather than a result of some underlying physics of the black hole background. Far from the horizon, the system is thus integrable, with periodic orbits and the appearance of unbroken tori. However, when both horizons approach near our integrable system, the near-horizon gravity makes the system highly nonlinear, leading to the breakdown of the regular KAM  tori and subsequent chaos. It is this onset and development of chaos due to the presence of both horizons that we wish to investigate in this work.

We thus consider the situation where a probe massless particle is subjected to two harmonic potentials   $\frac{1}{2}K_r(r-r_0)^2$ and $\frac{1}{2}K_\theta(z-z_0)^2$ along $r$ and $\theta$ directions, respectively. The terms $K_r$ and $K_\theta$ represent the oscillator strengths along $r$ and $\theta$ directions, respectively (we also introduce a new variable $z=r_{h}\theta$ while writing the dynamical equations). Here, $r_0$ and $z_0(=r_h\theta_{0})$ are the equilibrium positions of these two harmonic potentials.

\begin{eqnarray}
	E=&&\sqrt{g(r)\Big[g(r)p_r^2+\dfrac{p_{\theta}^2}{r^2}\Big]}+\frac{1}{2}K_r(r-r_0)^2 \nonumber\\ 
	&&+\frac{1}{2}K_\theta~r^2_{h}(\theta-\theta_0)^2.\label{a2_1}
\end{eqnarray}
Let us note that the radial momenta $p_{r}$ and cross radial momenta $p_{\theta}$, which appear in the above energy expression, are the usual canonical momenta that can be derived from the Lagrangian of a massless particle in the Schwarzwald coordinate system.

Correspondingly, the dynamical equations of motion have the following form:
\begin{eqnarray}
	\dot{r}&=&\dfrac{\partial E}{\partial p_r}=\frac{g^2(r)p_r}{\sqrt{g^2(r)p_r^2+\dfrac{g(r)p_{\theta}^2}{r^2}}},\label{a3}
	\\
	\dot{p_r}&=&-\frac{\partial E}{\partial r}=-\dfrac{g(r)g'(r)p_r^2}{\sqrt{g^2(r)p_r^2+\dfrac{g(r)p_{\theta}^2}{r^2}}}-K_r(r-r_0)\nonumber\\
	&&-\dfrac{g'(r)p_{\theta}^2}{2r^2\sqrt{g^2(r)p_r^2+\dfrac{g(r)p_{\theta}^2}{r^2}}}+\dfrac{g(r)p_{\theta}^2}{r^3\sqrt{g^2(r)p_r^2+\dfrac{g(r)p_{\theta}^2}{r^2}}},\nonumber\\
	\label{a4}
	\\
	\dot{\theta}&=&\frac{\partial E}{\partial p_{\theta}}=\dfrac{g(r)p_{\theta}}{r^2\sqrt{g^2(r)p_r^2+\dfrac{g(r)p_{\theta}^2}{r^2}}},\label{a5}
	\\
	\dot{p_{\theta}}&=&-\frac{\partial E}{\partial\theta}=-K_{\theta}~r^2_h(\theta-\theta_0).\label{a6}
\end{eqnarray}
where the derivative is taken with respect to an affine parameter $\tau$. These are the main equations for the numerical studies to be performed.

\section{Numerical Analysis} \label{sec4}
\subsection{Analysis of Poincar$\Acute{e}$ sections}
In this section, we will examine the role played by the black hole horizon and cosmological horizon towards the onset of chaos (i.e., the first appearance of broken tori in the associated Poincar$\Acute{e}$ section) for our massless particle confined in a harmonic potential.

The Poincar$\Acute{e}$ map, an essential tool for studying nonlinear dynamics, is defined as the intersection of periodic/ aperiodic orbits with a subspace transverse to the trajectories residing in the full state space. The essential idea is to map the higher dimensional phase-space trajectories into the lower one using the Poincar$\Acute{e}$ map. For this model setup, we choose the $\theta=0$ plane as the Poincar$\Acute{e}$ section. Within this space, we plot the points on the $(r-p_r)$ plane when the particle intersects the Poincar$\Acute{e}$ section, guided by the constraint of fixed energy $E$ and $p_{\theta}>0$. For the periodic case, the plotted points will lie on a torus in the phase space, while for the chaotic scenario, some of these tori will be broken. Such broken tori in different corners of the phase space is a well-recognized signature of chaos.

Now, we will numerically solve the dynamical equations of motion \Big(Eq. \eqref{a3}, Eq. \eqref{a4}, Eq. \eqref{a5}, and Eq. \eqref{a6}\Big) for the SdS black hole background. These numerical solutions will aid in the construction of the Poincar$\Acute{e}$ section.

As the model demands the range of $y$ from $0$ to $1/27 (\approx 0.037)$, we will focus on the chaotic dynamics in this range. We solve the dynamical equations of motion \Big(Eq. \eqref{a3}, Eq. \eqref{a4}, Eq. \eqref{a5} and Eq. \eqref{a6}\Big) using the fourth order Runge-Kutta method with fixed step size $h=0.01$. In this analysis, we have chosen $K_r=100$, $K_{\theta}=25$ and $\theta_0=0$. We set the equilibrium position of the harmonic oscillator $r_0$ to $4.3$, ensuring that our integrable system stays in between the two horizons (in order to see how these two horizons affect form both sides). The other free variables $r,~p_r$ and $\theta$ are initialized randomly within the range $3.0<r<6.5~,-0.5<p_r<0.5$ and $-0.05<\theta<0.05$, respectively. The value of $p_{\theta}$ is obtained from Eq. \eqref{a2} for a fixed value of the conserved energy $E$ for the total system. The different colors in the following figures denote trajectories of the particle solved for those randomly chosen initial conditions.

\begin{figure}[H]
	\centering
	\includegraphics[width=0.8\linewidth,height=0.6\linewidth]{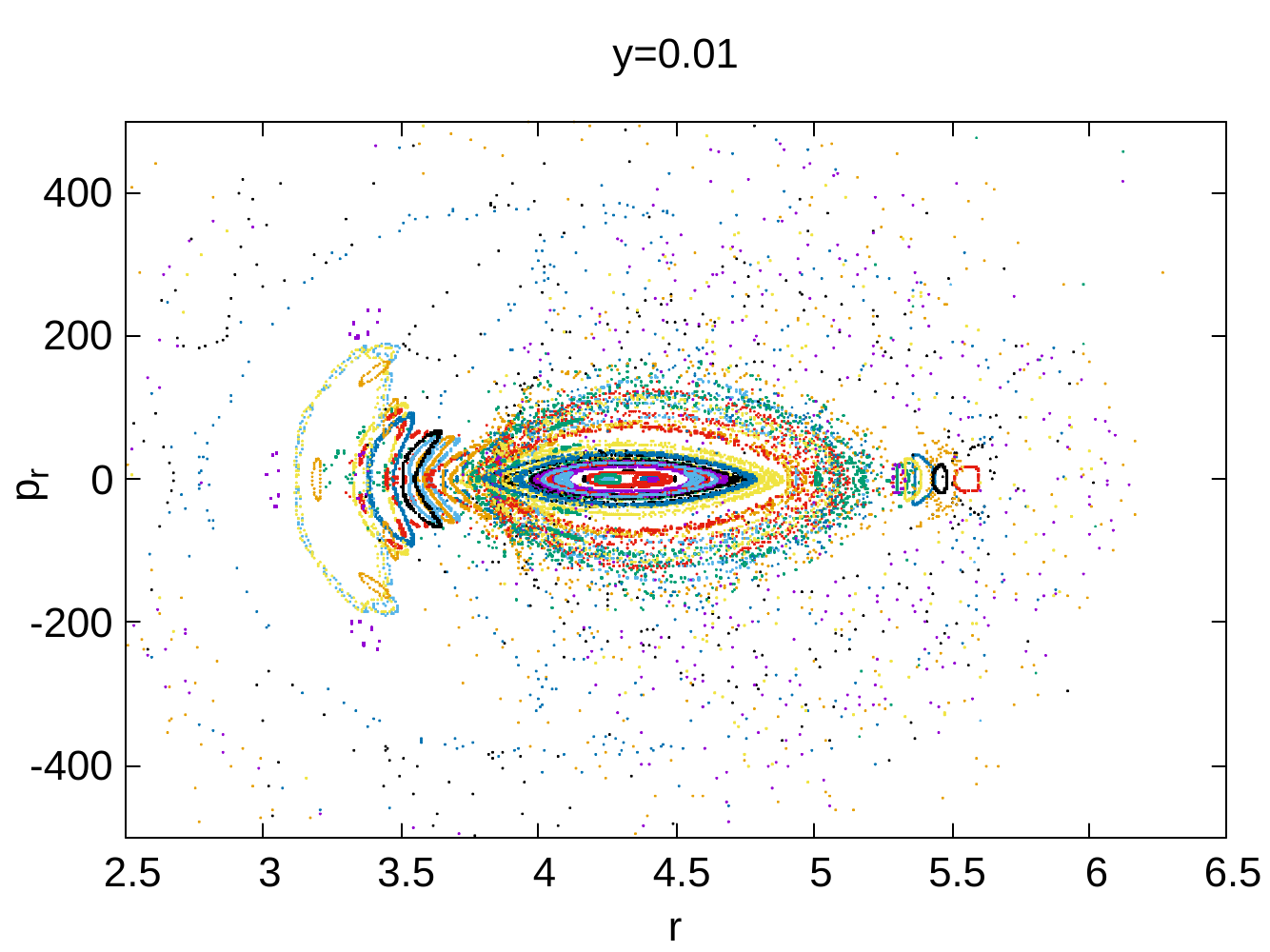}\label{1b}
	\includegraphics[width=0.8\linewidth,height=0.6\linewidth]{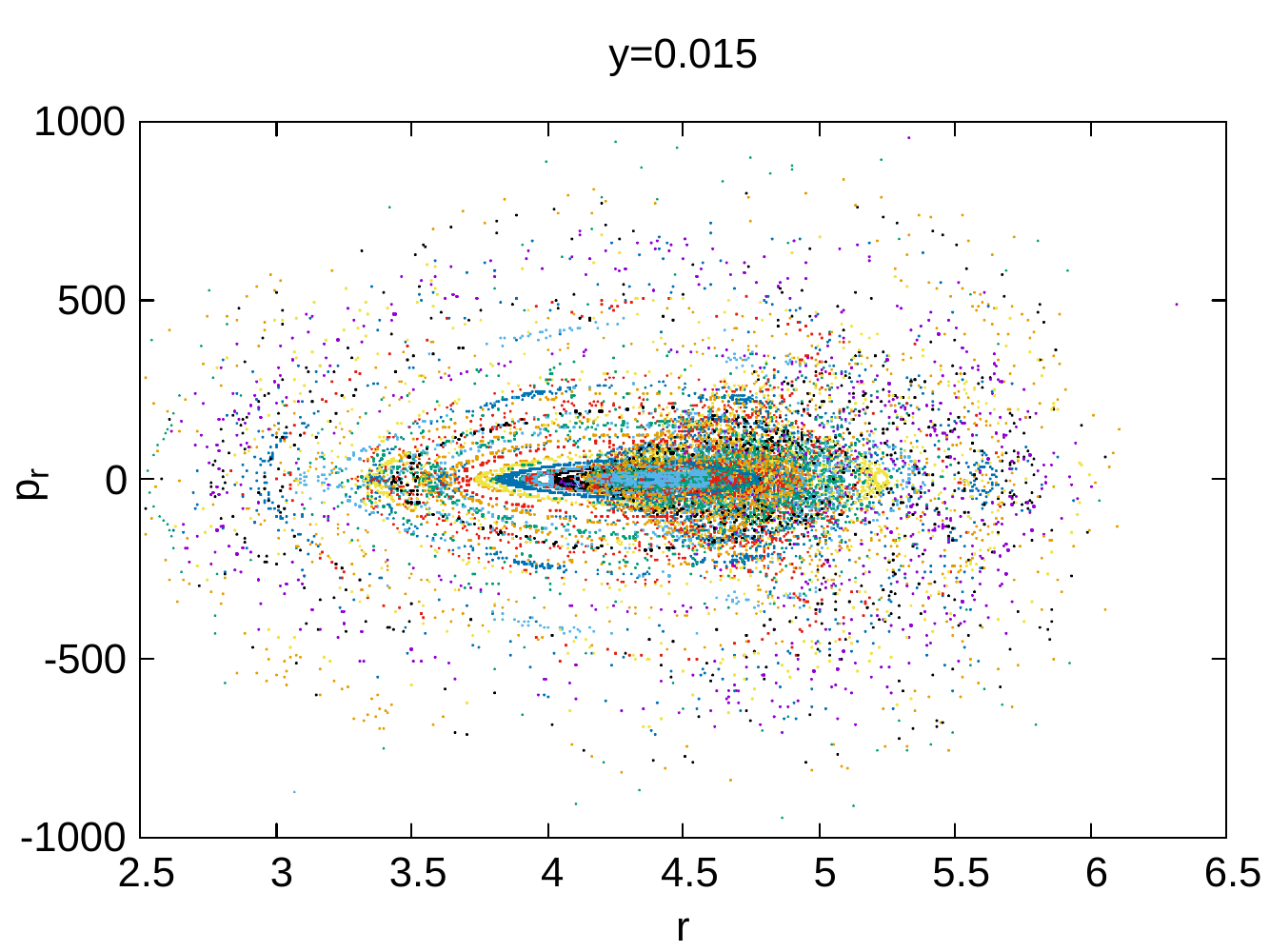}\label{1c}
	\includegraphics[width=0.8\linewidth,height=0.6\linewidth]{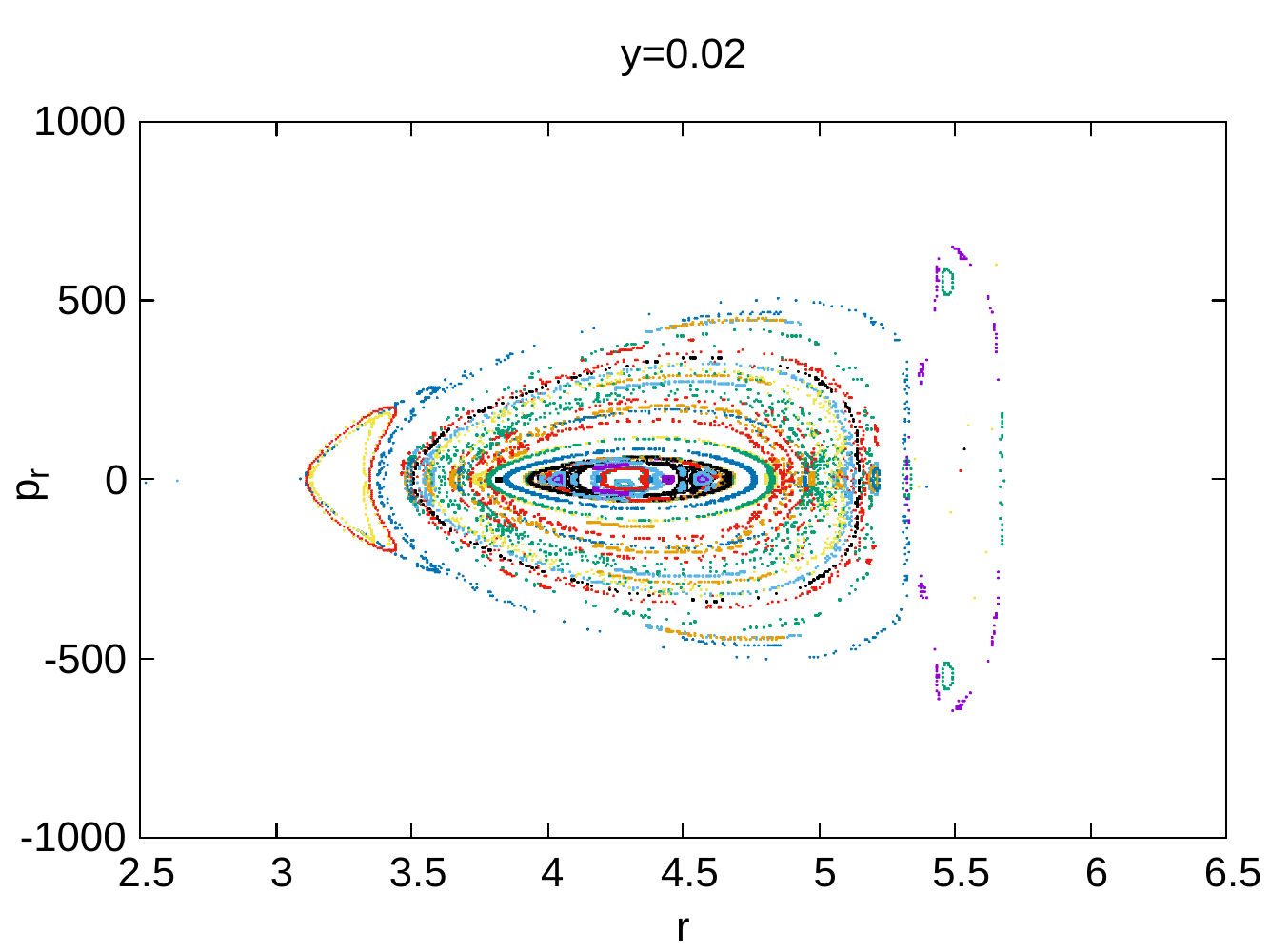}\label{1d}
	\includegraphics[width=0.8\linewidth,height=0.6\linewidth]{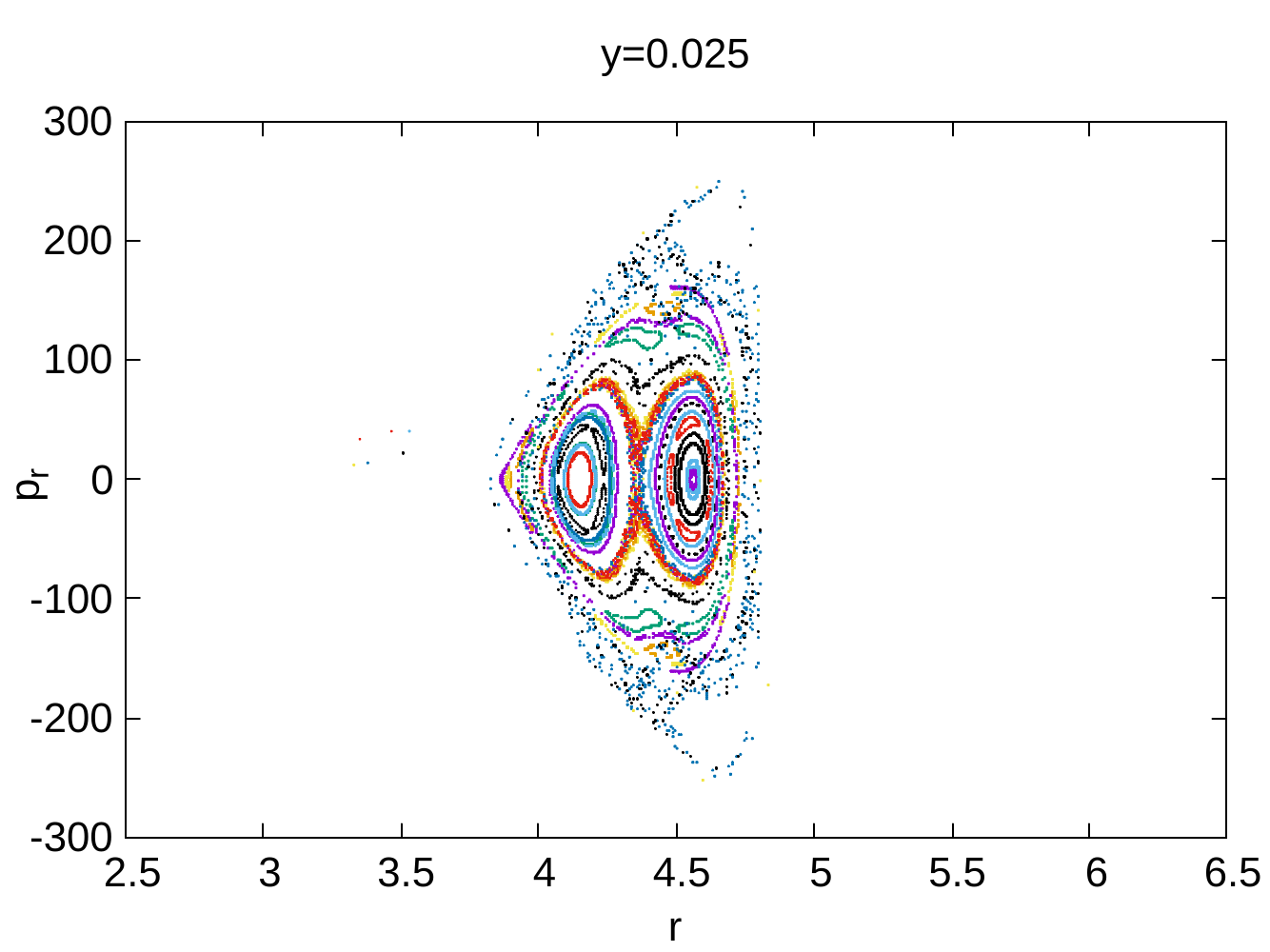}\label{1e}
	\caption{In these plots, we illustrate the Poincar$\Acute{e}$ section of a particle's trajectory projected onto the $(r-p_r)$ phase plane, setting $\theta=0$  with $p_{\theta}>0$, considering five different values of $y$: $y=0.01, 0.015, 0.02$, and $0.025$. At the lowest value of $y=0.01$, the Poincaré sections display regular Kolmogorov-Arnold-Moser (KAM) tori, indicating that the orbit is predominantly confined near the center of the harmonic potential, which is taken as $r_0=4.3$. However, as the value of $y$ increases, the position of the two horizons ($r_h$ and $r_c$) gradually approaches near to our integrable system. As a result of the Poincar$\Acute{e}$ sections start featuring distorted tori from both sides indicative of chaotic behavior.}
	\label{plot2}
\end{figure}

In these plots, we present the Poincaré sections of a particle's trajectory, projected onto the \((r-p_r)\) phase plane with \(\theta=0\) and \(p_{\theta} > 0\), for five distinct values of \(y\): \(y = 0.01, 0.015, 0.02\), and \(0.025\). At the smallest value, \(y = 0.01\), the Poincaré sections exhibit regular Kolmogorov-Arnold-Moser (KAM) tori, indicating that the particle's motion is primarily confined near the center of the harmonic potential at \(r_0 = 4.3\).  

As \(y\) increases, the locations of the two horizons, \(r_h\) (black hole horizon) and \(r_c\) (cosmological horizon), move closer to the region of the integrable system, consistent with Fig. (\ref{plot2}). Consequently, the Poincaré sections begin to show distorted tori on both sides, signaling the onset of chaotic behavior. For \(y = 0.025\), the outer portion of the Poincaré section is entirely engulfed by the horizon. Additionally, the cosmological horizon (\(r_c\)) becomes visible at the right edge of the plot, as indicated by the absence of data points beyond \(r = 4.8\).  

Further increasing \(y\) causes the Poincaré sections to contract as the two horizons approach each other. At \(y = 0.37\), the Poincaré section vanishes altogether, signifying the transition to a Nariai spacetime, as discussed earlier.

Now, let us discuss each plot of Fig. (\ref{plot2}) in a more detailed way. At this value of \( y = 0.01 \), the event horizon is located at \( r_h = 2.09 \), while the cosmological horizon is farther away at \( r_c = 8.73 \). The Poincaré section predominantly displays regular Kolmogorov-Arnold-Moser (KAM) tori, especially near the center of the harmonic potential at \( r_{0} = 4.3 \). However, on the left-hand side, closer to the event horizon, broken tori and scattered points are more prominent. This indicates that the event horizon's proximity has already begun to destabilize the system and induce chaos. On the other hand, the right-hand side, being farther from the cosmological horizon, retains a more regular structure as the influence of \( r_c \) has not yet become significant.

At \( y = 0.015 \), the horizons move closer to the system: \( r_h = 2.15 \) and \( r_c = 6.84 \). The Poincaré section shows a marked increase in scattered points on the left-hand side, reflecting the stronger influence of the event horizon. The system here is predominantly dominated by the event horizon’s effect, driving significant chaotic behavior in the left-hand region. On the right-hand side, near \( r_c \), some initial signs of broken tori and scattered points emerge, signaling the beginning of the cosmological horizon's influence.

At \( y = 0.02 \), the horizons have moved even closer: \( r_h = 2.21 \) and \( r_c = 5.67 \). The event horizon's proximity leads to some phase points being engulfed entirely, particularly on the left-hand side. This effect suppresses regular motion in the vicinity of \( r_H \). Meanwhile, the right-hand side begins to exhibit more pronounced chaotic behavior due to the cosmological horizon, as evidenced by the scattered points and fragmented tori.

At \( y = 0.025 \), the horizons approach very close to the system, with \( r_h = 2.3 \) and \( r_c = 4.83 \). Both horizons exert strong influences, squeezing the accessible phase space and leaving the Poincaré section compressed near the center. Most phase points on both the left and right sides are now engulfed by the horizons. The chaotic features dominate, and the previously distinguishable regions of regular motion have largely disappeared.

These observations highlight the inevitable emergence of chaos due to the horizons. However, the effects of the event horizon are evident from the very beginning, as it remains closer to the system across all values of \( y \). The cosmological horizon, initially distant, begins to contribute significantly to the chaos as it approaches the system with increasing \( y \). This interplay between the two horizons gradually destabilizes the system, transitioning it from predominantly regular to chaotic dynamics. The final plot (\( y = 0.025 \)) encapsulates the culmination of these effects, with the combined influence of the horizons dominating the dynamics entirely.

Therefore, it is evident that the chaotic influence of the event horizon on the system manifests earlier and more prominently than that of the cosmological horizon. This is because the event horizon is closer to the system from the outset, allowing it to destabilize the particle motion and disrupt regular trajectories in the phase space. In contrast, the cosmological horizon’s influence becomes significant only when it approaches closer to the system. \textit{This difference suggests that, within the parameter range considered, the event horizon exerts a relatively stronger chaotic influence on the system due to its consistent proximity.}

\subsection{Analysis of Lyapunov Exponents}
We now proceed to investigate the Lyapunov exponents ($\lambda_{L}$) to quantify the chaos observed in the Poincar$\Acute{e}$ sections studied in the previous section. The Lyapunov exponent of a dynamical system is defined as the quantity that characterizes the rate of separation of two infinitesimally close trajectories \cite{Sandri}. On the other hand, the value of the maximum Lyapunov exponent in this context is defined as \cite{Sandri, Strogatz}
\begin{equation}
	\lambda_{L,max}=\lim_{\tau\to\infty}\frac{1}{\tau}\ln\Big(\frac{\delta r(\tau)}{\delta r(0)}\Big).\label{LYP_max}
\end{equation} 
The term $\delta r(\tau)$ in the numerator represents the separation between two initially close trajectories at time $\tau$, and $\delta r(0)$ in the denominator represents the separation at the initial time.  Since we are considering null geodesics, Eq. \eqref{LYP_max} implicitly defines the affine parameter $\tau$.

Let us note that in this classical setting, there is an upper bound on $\lambda_{L}$ given by the surface gravity ($\kappa$) of the black hole \cite{Hashimoto_2017}, which is related through the temperature of the system as well introduced by  Maldacena-Shenker-Stanford (MSS) \cite{Maldacena_2016}
\begin{eqnarray}
	\lambda_L \leq \frac{2\pi T}{\hbar}
\end{eqnarray}
where, $\lambda_L$ is the largest Lyapunov exponent of the system and $T$ is the temperature of the system. Therefore, in this scenario our main motivation here is to see whether this bound is satisfied for our system or not when both the horizons are approaching near our system. 
\begin{figure}[H]
	\begin{center}
		\includegraphics[width=1.0\linewidth]{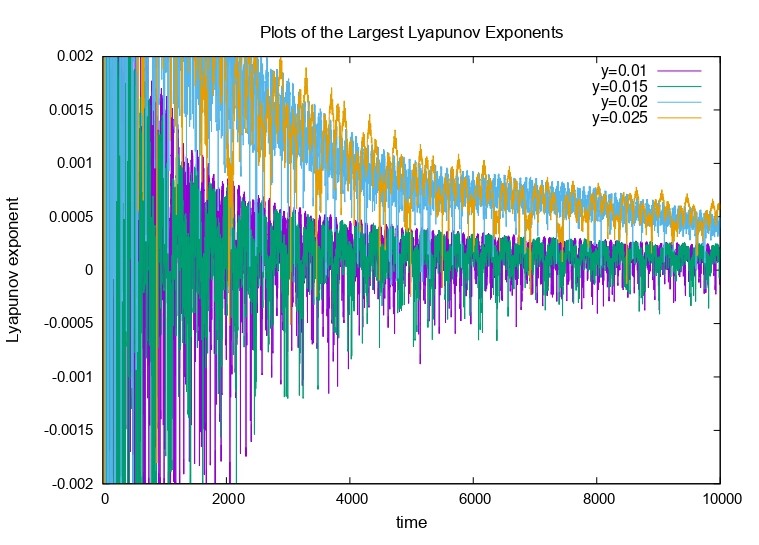}\label{plot4}
	\end{center}
	\caption{Plots of the largest Lyapunov exponents for different values of $y$ for a constant value of the total energy of the system $E=10$.}
	\label{fig3}
\end{figure}
\noindent Now, before going into the analysis of the Lyapunov exponent, let us discuss briefly how to compute effective temperature for a two-horizon scenario. The surface gravity of the event horizon and the cosmological horizon is given respectively as follows \cite{Bousso_1996, Bousso_1998, Shankaranarayanan_2003, Choudhury:2004ph}
\begin{eqnarray}
	\kappa_h &=& \alpha \Big|\frac{M}{r_h^2} - \frac{r_h}{l^2}\Big|, \\
	\kappa_c &=& \alpha \Big|\frac{M}{r_c^2} - \frac{r_c}{l^2}\Big|.
\end{eqnarray}
From the relations given in eqns (\ref{rh}), (\ref{rc}) and (\ref{psi}) we can compute $r_h$ and $r_c$. The parameter $\alpha$ is defined as
\begin{equation}
	\alpha = \frac{1}{\sqrt{1-(27y)^{1/3}}}.
\end{equation}
The ranges of $\kappa_h$ and $\kappa_c$ are given below
\begin{eqnarray}
	&&\frac{l}{\sqrt{3}} < \frac{1}{\kappa_h} < \frac{4l}{3\sqrt{3}}, \\
	&&\frac{l}{\sqrt{3}} < \frac{1}{\kappa_c} < l \, \, .
\end{eqnarray}
As shown in \cite{Shankaranarayanan_2003} we can define an `effective' surface gravity, which we call from now on $\kappa_{e}$, for the event horizon and the cosmological horizon as 
\begin{equation}
	\kappa_e = \Big( \frac{1}{\kappa_h} + \frac{1}{\kappa_c} \Big)^{-1} .
\end{equation}
The range of $\kappa_e$ is
\begin{equation}
	\frac{1}{l} < \kappa_e < \sqrt{3} l \, \, .
\end{equation}
The effective temperature is given by,
\begin{equation}
	T_e = \frac{\kappa_e}{2 \pi} \, \, .
\end{equation}
Now, in Fig. (\ref{fig3}), we present the largest Lyapunov exponents for four different values of \( y \) while keeping the total energy of the system fixed at \( E=10 \), with all other parameters remaining unchanged. As evident from Fig. (\ref{fig3}), increasing \( y \) from 0.01 to 0.025 leads to a corresponding increase in the largest Lyapunov exponents, indicating enhanced chaotic behavior. This trend is expected due to the influence of both horizons in close proximity to the system. Furthermore, examining the saturation values of the Lyapunov exponents over longer time scales, we find that they consistently adhere to the MSS bound. The saturation values of the Lyapunov exponent and the corresponding system temperature are summarized in the following table:


	\begin{table*}[htbp]
		\centering
		\renewcommand{\arraystretch}{1.5} 
		\begin{tabular}{|p{0.25\linewidth} |p{0.35\linewidth} |p{0.35\linewidth}|}
			\toprule
			
			\rowcolor{mymaroon}
			\multicolumn{1}{|c|}{\textbf{\textcolor{white}{\centering{Value of $y$}}}} & \textbf{\textcolor{white}{The saturated value of the Lyapunov exponent ($\lambda_{L,max}$) from Fig. (\ref{fig3})}} & \multicolumn{1}{|c|}{\textbf{\textcolor{white}{Temperature ($T_e$)}}} \\
			\midrule

			\cellcolor{lightblue} 0.01 & \cellcolor{lightgreen} $\approx$ 0.000105& \cellcolor{masteredyellow} 0.0236749 \\
			
			\cellcolor{lightblue} 0.015 & \cellcolor{lightgreen} $\approx$ 0.000145& \cellcolor{masteredyellow} 0.0325972 \\
			
			\cellcolor{lightblue} 0.02 & \cellcolor{lightgreen} $\approx$ 0.000487 & \cellcolor{masteredyellow} 0.0430485 \\
			
			\cellcolor{lightblue} 0.025 & \cellcolor{lightgreen} $\approx$ 0.000575 & \cellcolor{masteredyellow} 0.0572411 \\

			\bottomrule
		\end{tabular}
		
		\caption{Comparison between the saturated value of the Lyapunov exponent ($\lambda_{L, max}$) from Fig. (\ref{fig3}) and the corresponding temperature of the system for different values of $y$. We observe that the largest Lyapunov exponent values respect the MSS bound.}
		\label{tab:table}
	\end{table*}
\section{Conclusion} \label{conclusion}
In this paper, we have analyzed the motion of a massless, chargeless particle in SdS spacetime. Our study reveals that the radial motion exhibits exponential growth, suggesting the potential onset of chaos in an otherwise integrable system influenced by the presence of horizons. This behavior is further confirmed by examining the Poincar$\Acute{e}$ sections of the trajectories, facilitated by introducing a harmonic trap to confine the particle’s motion.

We explore how the black hole and cosmological horizons affect the particle’s trajectories when the system is positioned between them.  When the horizons are far apart, the Poincar$\Acute{e}$ sections display regular Kolmogorov-Arnold-Moser (KAM) tori, indicating that the particle’s motion remains predominantly confined near the center of the harmonic potential. However, as the horizons approach the integrable region, the Poincar$\Acute{e}$ sections exhibit increasingly distorted tori on both sides, signaling the onset of chaotic behavior. When the horizons coincide, the Poincar$\Acute{e}$ sections collapse entirely, vanishing as the system transitions to Nariai spacetime.

Additionally, our analysis reveals that the chaotic influence of the event horizon on the system manifests earlier and more prominently than that of the cosmological horizon. This is because the event horizon is closer to the system from the outset, allowing it to destabilize the particle motion and disrupt regular trajectories in the phase space. In contrast, the cosmological horizon’s influence becomes significant only when it approaches closer to the system. This difference suggests that, within the parameter range considered, the event horizon exerts a relatively stronger chaotic influence on the system due to its consistent proximity.

To quantify the observed chaos, we have also examined the Lyapunov exponents. Our results show that as the parameter $y$ increases from 0.01 to 0.025, the largest Lyapunov exponent grows correspondingly, indicating enhanced chaotic behavior. This trend aligns with expectations, given the stronger influence of closely spaced horizons on the system. Furthermore, by analyzing the saturation values of the Lyapunov exponents over longer timescales, we found that they consistently adhere to the MSS bound.

The findings presented in this work highlight the profound role played by horizons in governing the chaotic dynamics of particles in curved spacetime. Our results suggest that horizons not only act as boundaries of causality but also serve as catalysts for chaotic behavior, affecting the motion of emitted particles. This insight raises several intriguing questions regarding the broader implications of chaos in black hole and cosmological settings. One promising direction for future investigation is to explore whether the chaotic dynamics of emitted particles can leave observational signatures in astrophysical settings. Given that particles escaping the black hole horizon may exhibit chaotic motion before reaching the cosmological horizon, their energy distribution and correlation properties could be affected in measurable ways. Such effects might manifest in cosmic ray spectra or gravitational wave echoes from near-horizon regions \cite{Cardoso:2016oxy, Oshita:2020dox}. Additionally, it would be interesting to extend this study to include quantum effects, particularly in the context of semiclassical gravity. The chaotic behavior observed here suggests that black hole horizons may influence the scrambling of information in ways that go beyond the classical picture. Investigating how quantum chaos indicators, such as out-of-time-ordered correlators (OTOCs) \cite{Maldacena_2016, Hashimoto:2017oit}, behave in SdS spacetime could provide deeper insights into the interplay between chaos and quantum information in black hole physics. Another avenue worth exploring is the potential connection between chaos in SdS spacetime and the dynamics of the early Universe. Since the cosmological horizon in our setup plays an analogous role to the de Sitter horizon in inflationary cosmology, studying chaotic behavior in this context might offer new perspectives on how initial perturbations evolved during cosmic inflation. This could have implications for understanding the nature of primordial fluctuations and their imprints on the cosmic microwave background. Finally, given the close parallels between our findings and other chaotic systems in modified gravity and higher-dimensional black hole spacetimes, a natural extension would be to generalize this study to different gravitational theories. In particular, investigating chaos in the presence of additional fields, such as scalar or gauge fields, could shed light on how fundamental interactions modify the onset of chaotic behavior in gravitational systems.

\section*{Acknowledgements}
We would like to thank Bibhas Ranjan Majhi for discussions and suggestions. We would also like to thank Anjan Sarakar for various suggestions regarding the numerical calculations that have been carried out here. 

\bibliography{main.bib}

\providecommand{\href}[2]{#2}\begingroup\raggedright\begin{thebibliography}{10}

\bibitem{LIGOScientific:2016aoc}
{\bfseries LIGO Scientific, Virgo} Collaboration, B.~P. Abbott {\em et~al.},
  ``{Observation of Gravitational Waves from a Binary Black Hole Merger},''
  \href{http://dx.doi.org/10.1103/PhysRevLett.116.061102}{{\em Phys. Rev.
  Lett.} {\bfseries 116} no.~6, (2016) 061102},
  \href{http://arxiv.org/abs/1602.03837}{{\ttfamily arXiv:1602.03837 [gr-qc]}}.

\bibitem{LIGOScientific:2017vwq}
{\bfseries LIGO Scientific, Virgo} Collaboration, B.~P. Abbott {\em et~al.},
  ``{GW170817: Observation of Gravitational Waves from a Binary Neutron Star
  Inspiral},'' \href{http://dx.doi.org/10.1103/PhysRevLett.119.161101}{{\em
  Phys. Rev. Lett.} {\bfseries 119} no.~16, (2017) 161101},
  \href{http://arxiv.org/abs/1710.05832}{{\ttfamily arXiv:1710.05832 [gr-qc]}}.

\bibitem{EHT:1}
{\bfseries Event Horizon Telescope} Collaboration, K.~Akiyama {\em et~al.},
  ``{First M87 Event Horizon Telescope Results. I. The Shadow of the
  Supermassive Black Hole},''
  \href{http://dx.doi.org/10.3847/2041-8213/ab0ec7}{{\em Astrophys. J. Lett.}
  {\bfseries 875} (2019) L1}, \href{http://arxiv.org/abs/1906.11238}{{\ttfamily
  arXiv:1906.11238 [astro-ph.GA]}}.

\bibitem{EHT:2}
{\bfseries Event Horizon Telescope} Collaboration, K.~Akiyama {\em et~al.},
  ``{First M87 Event Horizon Telescope Results. II. Array and
  Instrumentation},'' \href{http://dx.doi.org/10.3847/2041-8213/ab0c96}{{\em
  Astrophys. J. Lett.} {\bfseries 875} no.~1, (2019) L2},
  \href{http://arxiv.org/abs/1906.11239}{{\ttfamily arXiv:1906.11239
  [astro-ph.IM]}}.

\bibitem{EHT:3}
{\bfseries Event Horizon Telescope} Collaboration, K.~Akiyama {\em et~al.},
  ``{First M87 Event Horizon Telescope Results. III. Data Processing and
  Calibration},'' \href{http://dx.doi.org/10.3847/2041-8213/ab0c57}{{\em
  Astrophys. J. Lett.} {\bfseries 875} no.~1, (2019) L3},
  \href{http://arxiv.org/abs/1906.11240}{{\ttfamily arXiv:1906.11240
  [astro-ph.GA]}}.

\bibitem{EHT:4}
{\bfseries Event Horizon Telescope} Collaboration, K.~Akiyama {\em et~al.},
  ``{First M87 Event Horizon Telescope Results. IV. Imaging the Central
  Supermassive Black Hole},''
  \href{http://dx.doi.org/10.3847/2041-8213/ab0e85}{{\em Astrophys. J. Lett.}
  {\bfseries 875} no.~1, (2019) L4},
  \href{http://arxiv.org/abs/1906.11241}{{\ttfamily arXiv:1906.11241
  [astro-ph.GA]}}.

\bibitem{EHT:5}
{\bfseries Event Horizon Telescope} Collaboration, K.~Akiyama {\em et~al.},
  ``{First M87 Event Horizon Telescope Results. V. Physical Origin of the
  Asymmetric Ring},'' \href{http://dx.doi.org/10.3847/2041-8213/ab0f43}{{\em
  Astrophys. J. Lett.} {\bfseries 875} no.~1, (2019) L5},
  \href{http://arxiv.org/abs/1906.11242}{{\ttfamily arXiv:1906.11242
  [astro-ph.GA]}}.

\bibitem{Maldacena_1999}
J.~Maldacena \href{http://dx.doi.org/10.1023/a:1026654312961}{{\em
  International Journal of Theoretical Physics} {\bfseries 38} no.~4, (1999)
  1113–1133}. \url{http://dx.doi.org/10.1023/A:1026654312961}.

\bibitem{Strominger_2001}
A.~Strominger, ``The ds/cft correspondence,''
  \href{http://dx.doi.org/10.1088/1126-6708/2001/10/034}{{\em Journal of High
  Energy Physics} {\bfseries 2001} no.~10, (Oct., 2001) 034–034}.
  \url{http://dx.doi.org/10.1088/1126-6708/2001/10/034}.

\bibitem{Dalui_2019}
S.~Dalui, B.~R. Majhi, and P.~Mishra, ``Presence of horizon makes particle
  motion chaotic,''
  \href{http://dx.doi.org/10.1016/j.physletb.2018.11.050}{{\em Physics Letters
  B} {\bfseries 788} (Jan., 2019) 486–493}.
  \url{http://dx.doi.org/10.1016/j.physletb.2018.11.050}.

\bibitem{Dalui_2020}
S.~Dalui, B.~R. Majhi, and P.~Mishra, ``Horizon induces instability locally and
  creates quantum thermality,''
  \href{http://dx.doi.org/10.1103/physrevd.102.044006}{{\em Physical Review D}
  {\bfseries 102} no.~4, (Aug., 2020) }.
  \url{http://dx.doi.org/10.1103/PhysRevD.102.044006}.

\bibitem{Dalui_2020_0}
S.~Dalui, B.~R. Majhi, and P.~Mishra, ``Induction of chaotic fluctuations in
  particle dynamics in a uniformly accelerated frame,''
  \href{http://dx.doi.org/10.1142/s0217751x20500815}{{\em International Journal
  of Modern Physics A} {\bfseries 35} no.~18, (June, 2020) 2050081}.
  \url{http://dx.doi.org/10.1142/S0217751X20500815}.

\bibitem{Dalui_2020_1}
S.~Dalui and B.~R. Majhi, ``Near-horizon local instability and quantum
  thermality,'' \href{http://dx.doi.org/10.1103/physrevd.102.124047}{{\em
  Physical Review D} {\bfseries 102} no.~12, (Dec., 2020) }.
  \url{http://dx.doi.org/10.1103/PhysRevD.102.124047}.

\bibitem{Dalui_2022}
S.~Dalui and B.~R. Majhi, ``Horizon thermalization of kerr black hole through
  local instability,''
  \href{http://dx.doi.org/10.1016/j.physletb.2022.136899}{{\em Physics Letters
  B} {\bfseries 826} (Mar., 2022) 136899}.
  \url{http://dx.doi.org/10.1016/j.physletb.2022.136899}.

\bibitem{Bera_2022}
A.~Bera, S.~Dalui, S.~Ghosh, and E.~C. Vagenas, ``Quantum corrections enhance
  chaos: Study of particle motion near a generalized schwarzschild black
  hole,'' \href{http://dx.doi.org/10.1016/j.physletb.2022.137033}{{\em Physics
  Letters B} {\bfseries 829} (June, 2022) 137033}.
  \url{http://dx.doi.org/10.1016/j.physletb.2022.137033}.

\bibitem{Das_2024}
S.~Das, S.~Dalui, and R.~Samanta, ``Near-horizon chaos beyond einstein
  gravity,'' \href{http://dx.doi.org/10.1103/physrevd.110.124037}{{\em Physical
  Review D} {\bfseries 110} no.~12, (Dec., 2024) }.
  \url{http://dx.doi.org/10.1103/PhysRevD.110.124037}.

\bibitem{Hashimoto_2017}
K.~Hashimoto and N.~Tanahashi, ``Universality in chaos of particle motion near
  black hole horizon,''
  \href{http://dx.doi.org/10.1103/physrevd.95.024007}{{\em Physical Review D}
  {\bfseries 95} no.~2, (Jan., 2017) }.
  \url{http://dx.doi.org/10.1103/PhysRevD.95.024007}.

\bibitem{C_ceres_2023}
E.~Cáceres, T.~Guglielmo, B.~Kent, and A.~Misobuchi, ``Out-of-time-order
  correlators and lyapunov exponents in sparse syk,''
  \href{http://dx.doi.org/10.1007/jhep11(2023)088}{{\em Journal of High Energy
  Physics} {\bfseries 2023} no.~11, (Nov., 2023) }.
  \url{http://dx.doi.org/10.1007/JHEP11(2023)088}.

\bibitem{Parikh_2000}
M.~K. Parikh and F.~Wilczek, ``Hawking radiation as tunneling,''
  \href{http://dx.doi.org/10.1103/physrevlett.85.5042}{{\em Physical Review
  Letters} {\bfseries 85} no.~24, (Dec., 2000) 5042–5045}.
  \url{http://dx.doi.org/10.1103/PhysRevLett.85.5042}.

\bibitem{Shankaranarayanan_2003}
S.~Shankaranarayanan, ``Temperature and entropy of schwarzschild–de sitter
  space-time,'' \href{http://dx.doi.org/10.1103/physrevd.67.084026}{{\em
  Physical Review D} {\bfseries 67} no.~8, (Apr., 2003) }.
  \url{http://dx.doi.org/10.1103/PhysRevD.67.084026}.

\bibitem{Sandri}
M.~Sandri, ``Numerical calculation of lyapunov exponents,'' {\em Math. J.}
  {\bfseries 6} (01, 1996) .

\bibitem{Strogatz}
S.~H. Strogatz, {\em Nonlinear dynamics and Chaos: with applications to
  physics, biology, chemistry, and engineering}.
\newblock Studies in nonlinearity. Addison-Wesley Pub, 1994.
\newblock
  \url{http://gen.lib.rus.ec/book/index.php?md5=fa4190d4e05f55134df61c889ad304b0}.

\bibitem{Maldacena_2016}
J.~Maldacena, S.~H. Shenker, and D.~Stanford, ``A bound on chaos,''
  \href{http://dx.doi.org/10.1007/jhep08(2016)106}{{\em Journal of High Energy
  Physics} {\bfseries 2016} no.~8, (Aug., 2016) }.
  \url{http://dx.doi.org/10.1007/JHEP08(2016)106}.

\bibitem{Bousso_1996}
R.~Bousso and S.~W. Hawking, ``Pair creation of black holes during inflation,''
  \href{http://dx.doi.org/10.1103/physrevd.54.6312}{{\em Physical Review D}
  {\bfseries 54} no.~10, (Nov., 1996) 6312–6322}.
  \url{http://dx.doi.org/10.1103/PhysRevD.54.6312}.

\bibitem{Bousso_1998}
R.~Bousso and S.~W. Hawking, ``(anti-)evaporation of schwarzschild–de sitter
  black holes,'' \href{http://dx.doi.org/10.1103/physrevd.57.2436}{{\em
  Physical Review D} {\bfseries 57} no.~4, (Feb., 1998) 2436–2442}.
  \url{http://dx.doi.org/10.1103/PhysRevD.57.2436}.

\bibitem{Choudhury:2004ph}
T.~R. Choudhury and T.~Padmanabhan, ``{Concept of temperature in multi-horizon
  spacetimes: Analysis of Schwarzschild-de Sitter metric},''
  \href{http://dx.doi.org/10.1007/s10714-007-0489-0}{{\em Gen. Rel. Grav.}
  {\bfseries 39} (2007) 1789--1811},
  \href{http://arxiv.org/abs/gr-qc/0404091}{{\ttfamily arXiv:gr-qc/0404091}}.

\bibitem{Cardoso:2016oxy}
V.~Cardoso, S.~Hopper, C.~F.~B. Macedo, C.~Palenzuela, and P.~Pani,
  ``{Gravitational-wave signatures of exotic compact objects and of quantum
  corrections at the horizon scale},''
  \href{http://dx.doi.org/10.1103/PhysRevD.94.084031}{{\em Phys. Rev. D}
  {\bfseries 94} no.~8, (2016) 084031},
  \href{http://arxiv.org/abs/1608.08637}{{\ttfamily arXiv:1608.08637 [gr-qc]}}.

\bibitem{Oshita:2020dox}
N.~Oshita, D.~Tsuna, and N.~Afshordi, ``{Quantum Black Hole Seismology I:
  Echoes, Ergospheres, and Spectra},''
  \href{http://dx.doi.org/10.1103/PhysRevD.102.024045}{{\em Phys. Rev. D}
  {\bfseries 102} no.~2, (2020) 024045},
  \href{http://arxiv.org/abs/2001.11642}{{\ttfamily arXiv:2001.11642 [gr-qc]}}.

\bibitem{Hashimoto:2017oit}
K.~Hashimoto, K.~Murata, and R.~Yoshii, ``{Out-of-time-order correlators in
  quantum mechanics},'' {\em JHEP} {\bfseries 10} (2017) 138,
  \href{http://arxiv.org/abs/1703.09435}{{\ttfamily arXiv:1703.09435
  [hep-th]}}.

\end{thebibliography}\endgroup

\bibliographystyle{utphys1}

\end{document}